\documentclass[runningheads]{llncs}
\usepackage[T1]{fontenc}
\usepackage{graphicx}
\usepackage[numbers]{natbib}
\usepackage{amssymb}
\usepackage{amsmath}
\usepackage{physics}
\usepackage{amsfonts}

\begin{document}
\title{Tuning Agent-Based Predator-Prey Models Toward Lotka-Volterra Dynamics}
\author{Corinna Mandl\inst{1}\orcidID{0009-0001-0198-1950}\thanks{These authors contributed equally.} \and
Siddharth Chaturvedi\inst{1}\orcidID{0009-0000-4965-1718}\textsuperscript{\thefootnote} \and
Marcel van Gerven\inst{1}\orcidID{0000-0002-2206-9098}}
\authorrunning{C. Mandl et al.}
\institute{Department of Machine Learning and Neural Computing\\Donders Institute for Brain, Cognition and Behaviour, Radboud University\\Nijmegen, the Netherlands\\ \email{siddharth.chaturvedi@donders.ru.nl}}

\maketitle              % typeset the header of the contribution
\begin{abstract}
Recent growth in compute power has made it increasingly feasible to use large-scale agent-based models to simulate complex adaptive systems. A central difficulty is that such models contain many local rules and parameters, where small changes can lead to runaway behaviour, population collapse, or saturation at artificial bounds. We study this problem in a continuous predator-prey system where sheep and wolves are active agents with local sensing, internal energy, and recurrent neural network-based controllers. We ask whether environmental and demographic parameters can be tuned so that the resulting population dynamics resemble classical Lotka-Volterra cycles. We optimise these parameters with a feature-based loss that rewards sustained oscillations, phase lag, bounded populations, and long-term persistence, first for random controllers and then for evolved controllers in a more naturalistic setting. The model is implemented in ABMax, a JAX-based agent-based modelling framework that enables efficient batched simulation on hardware accelerators.

\end{abstract}

\section{Introduction}

Agent-based models (ABMs) connect local interactions among individual agents with population-level dynamics, making them useful for studying complex adaptive systems in ecology, artificial life, neuroscience, and economics~\citep{holland1992complex,deangelis2005individual}. However, their flexibility also makes them difficult to control. Many ABMs contain nonlinear interactions whose aggregate outcomes can be multi-modal, skewed, or fat-tailed. In such settings, rare large events are more likely than under thin-tailed statistics, making model dynamics sensitive to local rules, initial conditions, and parameters~\citep{helbing2012agent,clauset2009power}. A central challenge is therefore to constrain ABM dynamics without removing the interactions that make them interesting.

Predator-prey systems provide a useful testbed for this problem. They are simple enough to analyse, but rich enough to produce sustained population-level structure through ecological feedback. Sheep support wolves, wolves suppress sheep, and the decline of one population changes the growth conditions of the other. This feedback loop is captured by the classical Lotka-Volterra (LV) equations~\citep{volterra1926fluctuations,idema2005behaviour}, which provide a principled macroscopic target for stabilising an agent-based predator-prey model.

Recent work has increasingly treated the relation between microscopic rules and macroscopic behaviour as a modelling problem in its own right. Macroscopic models can be derived from microscopic swarm descriptions and used to guide individual rules~\citep{quan2026macroscopic}, while ABMs can be calibrated from data by fitting parameters or latent micro variables to reproduce aggregate patterns~\citep{monti2023learning}. In predator-prey systems, ABMs have been reduced to coarse-grained stochastic dynamics~\citep{niemann2021datadriven}, and LV models and ABMs have been used as complementary descriptions of the same process~\citep{hodzic2016complex}. Our work follows this micro-macro tradition, but uses the desired macroscopic regime as an explicit top-down optimisation target. Rather than reducing an ABM to a coarse LV-like model, we tune the microscopic parameters of a continuous predator-prey ABM so that its population trajectories satisfy LV-like constraints.

We model sheep and wolves as self-propelled active agents moving in a continuous two-dimensional arena~\citep{yaya2023predator}. The interaction rules are inspired by the canonical Wolf Sheep predation model~\citep{Wilensky1997}. Agents also have a continuous-time recurrent neural network (CTRNN) controller to control their motion. The optimisation is carried out in two separate stages. First, we optimise agent controllers that determine sensory-motor behaviour. Second, we freeze either evolved or random controllers and optimise ABM-level ecological parameters controlling energy intake, predation, metabolism, birth, and death. The rest of the paper describes the model, optimisation procedure, results, and limitations.

\section{Model}\label{sec:model}

\begin{figure}
\centering
\includegraphics[width=1.0 \textwidth]{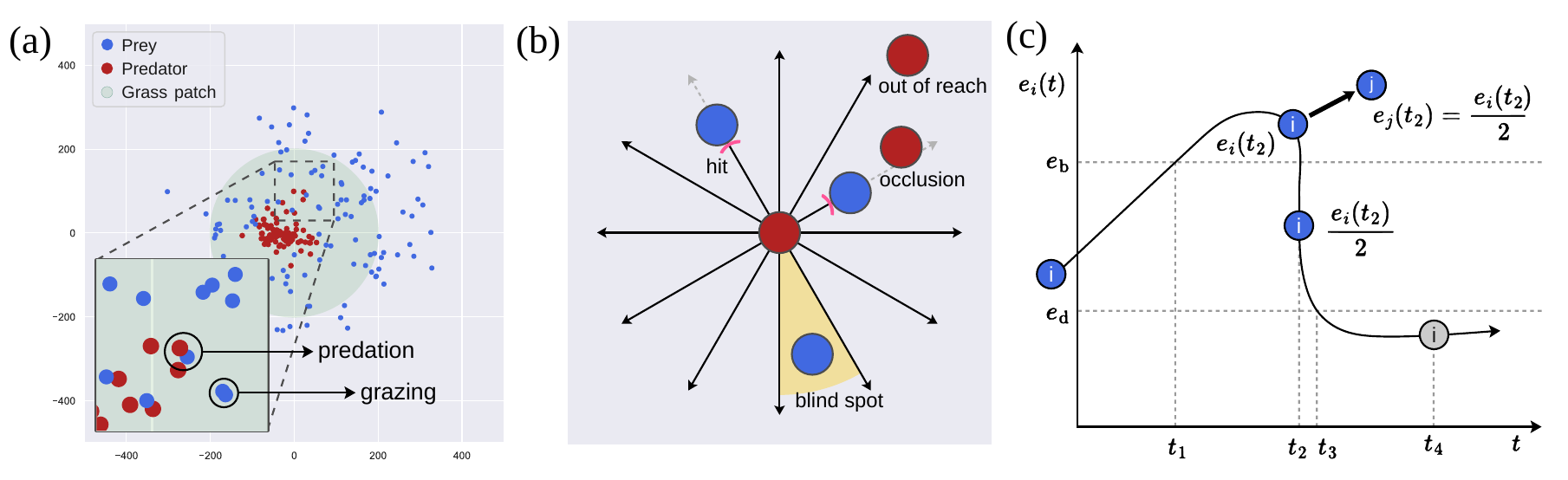}
\caption{\textbf{(a)} Snapshot of model with agent interactions. \textbf{(b)} Different sensing scenarios. \textbf{(c)} Energy during the birth-death process of agents. } \label{fig1}
\end{figure}

The model consists of circular active agents moving in a two-dimensional Cartesian plane (Fig.~\ref{fig1}a). It is inspired by related \texttt{ABMax} models of active particles and ecological agent-based systems~\citep{chaturvedi2025emergence,chaturvedi2026role}. Agents do not undergo hard-body collisions and can pass through each other. The system is integrated with a forward Euler scheme of step size $\Delta t$, with time evaluated on $t\in \{k \Delta t \colon k=0,1,\dots, T\}$.

\subsection{Population turnover}\label{subsec:population_turnover}

The agent population $\mathcal{A}=\mathcal{W}\, \cup \, \mathcal{S}$ is divided into sheep $\mathcal{S}$ and wolves $\mathcal{W}$, representing prey and predators. To preserve static array shapes during JAX compilation, each species is represented by a fixed-size buffer. The model accommodates at most $N^{\mathcal{S}}_{\max}$ sheep and $N^{\mathcal{W}}_{\max}$ wolves, of which $N^{\mathcal{S}}(t)$ and $N^{\mathcal{W}}(t)$ are active at time $t$.

Population turnover follows threshold-based birth and death. Inspired by the canonical Wolf Sheep model~\citep{Wilensky1997}, an agent $i \in \mathcal{A}$ is eliminated when its internal energy $e_i(t)$ remains below a death threshold $e_d$ for an uninterrupted interval $t_d$. In contrast to stochastic reproduction in the canonical model, an active agent becomes viable for reproduction when $e_i(t)$ remains above a birth threshold $e_b$ for an uninterrupted interval $t_b$. Reproduction occurs only if an inactive slot of the same species is available. The offspring inherits the parent's controller parameters, receives half of the parent's energy, and is initialised near the parent by perturbing its Cartesian coordinates within $[-d_s,d_s]$ along each axis. The parent energy is also halved, introducing a natural delay before the next birth (Fig.~\ref{fig1}c).
% Table: $N^{(\mathcal{W})}_{\max}$,  $N^{(\mathcal{S})}_{\max}$, $e_d$ $e_b$, $t_d$ $t_b$
% table: init conditions: $N^{(\mathcal{W})}(0)$ N^{(\mathcal{S})}(0)$

\subsection{Position model}

The position of the centre of an active agent is $\mathbf{q}_i(t)\in\mathbb{R}^2$, and its orientation is $\theta_i(t)\in[-\pi,\pi)$. Positions are clipped component-wise to the rectangular arena. Dropping the agent index for readability, the agent positions are updated as
\begin{align}
    \dot{\mathbf{q}}(t) &= \mathbf{v}(t), 
    \qquad 
    \mathbf{v}(t) = s(t)
    \begin{bmatrix}
        \cos\theta(t) \\
        \sin\theta(t)
    \end{bmatrix}, \label{eq:position_model} \\
    \dot{\theta}(t) &= \omega(t),
    \qquad
    \omega(t) = u(t). \label{eq:orientation_model}
\end{align}
Here, \(\mathbf{v}(t)\in\mathbb{R}^2\) is the Cartesian velocity and
\(\omega(t)\in\mathbb{R}\) is the angular velocity. The translational and
rotational velocity commands \(s_i(t)\) and \(u_i(t)\) for an agent $i$ are obtained from the
controller readouts as
\begin{align}
    s_i(t) &= k^{(s)} \tanh\!\left(a_i^{(1)}(t)\right)
    \left(1+\epsilon \xi_i^{(s)}(t)\right), \label{eq:linear_speed} \\
    u_i(t) &= k^{(u)}\tanh\!\left(a_i^{(2)}(t)\right)
    \left(1+\epsilon \xi_i^{(u)}(t)\right), \label{eq:angular_speed}
\end{align}
where $\epsilon$ is a positive scalar by which Gaussian noise terms $\xi^{(s)}(t) , \xi^{(u)}(t)\sim \mathcal{N}(0,1)$ are multiplied. The translational and angular speeds are scaled by positive constants $k^{(s)}$ and $k^{(u)}$, respectively. The quantities $a_i^{(1)}(t),a_i^{(2)}(t)\in\mathbb{R}$ are the two raw readouts of the continuous-time recurrent neural network controller.

%Table: init conditions: $\mathbf{q}}(0)$, $\mathbf{v}(0)$, $\theta(0)$ $\omega(0)$
%Table: $k^{(s)}$, $k^{(u)}$, $\epsilon$

\subsection{Energy model}

Each active agent maintains an internal energy $e_i(t)$. Sheep gain energy from a circular grass patch, wolves gain energy by consuming sheep, and both species lose a constant metabolic cost. For sheep $i\in\mathcal{S}(t)$, grass intake is
\begin{equation}
    e^{(G)}_i(t) = k^{(G)} \delta^{(G)}(\mathbf{q}_i(t)),
\end{equation}
where $k^{(G)}$ is the grass intake rate and $\delta^{(G)}(\cdot)$ is $1$ when the sheep lies within the grass patch of radius $d_G$ centred at the origin, and $0$ otherwise.

Predation is a local energy transfer from sheep to wolves. Each wolf consumes at most one sheep at a time, namely the closest active sheep within its capture distance $d_{\mathcal{A}}$. We define the capture indicator as a Boolean-step function given by
$\delta^{(P)}_{j i}(t)$ as
\begin{equation}
    \delta^{(P)}_{j i}(t)=
    \begin{cases}
        1, & \text{if } i\in\mathcal{S}(t) \text{ is nearest to } j\in \mathcal{W}(t)
        \text{ and } \|\mathbf{q}_i(t)-\mathbf{q}_j(t)\|_2 \le d_{\mathcal{A}},\\
        0, & \text{otherwise.}
    \end{cases}
\end{equation}
When sheep $i$ is caught by wolf $j$, a fraction $k^{(P)}$ of the sheep's
current energy is transferred. Thus, the predation loss of sheep $i$ is
\begin{equation}
    e^{(P-)}_i(t) =
    k^{(P)} e_i(t)\sum_{j\in\mathcal{W}(t)} \delta^{(P)}_{j i}(t),
\end{equation}
and the predation gain of wolf \(j\) is
\begin{equation}
    e^{(P+)}_j(t) =
    k^{(P)} \sum_{i\in\mathcal{S}(t)} e_i(t)\delta^{(P)}_{j i}(t).
\end{equation}
This allows multiple wolves to target the same sheep, in which case the sheep
loses the corresponding energy fraction once for each wolf that catches it.
The energy updates for sheep and wolves become
\begin{align}
    e_i(t+\Delta t) &= e_i(t) + e^{(G)}_i(t) - e^{(P-)}_i(t) - \mu^{(\mathcal{S})},
    \qquad i\in\mathcal{S}(t), \label{eq:sheep_energy} \\
    e_i(t+\Delta t) &= e_i(t) + e^{(P+)}_i(t) - \mu^{(\mathcal{W})},
    \qquad i\in\mathcal{W}(t). \label{eq:wolf_energy}
\end{align}
The internal energy of agents is further clipped between $e_{\min}$ and $e_{\max}$. If an agent's energy remains below the death threshold for the required duration, it is removed from the active
population. If it remains above the reproduction threshold for a certain duration, and a free slot is available, it reproduces, as described in Section~\ref{subsec:population_turnover}.

\subsection{Lotka-Volterra objective}\label{subsec:lv_objective}

In this subsection, we explain the empirically engineered loss function that is used to tune the discrete population dynamics of the predation ABM according to smooth Lotka-Volterra dynamics. The chief ingredient of the loss function is the Pearson correlation function~\citep{lee1988thirteen}. For two equal-length time series $a(t)$ and $b(t)$, the Pearson correlation is defined as
\begin{equation}
    C(a,b)=
    \frac{
        \left\langle (a(t)-\langle a\rangle)(b(t)-\langle b\rangle)\right\rangle
    }{
        \left(\sigma^{(a)}+\epsilon^{(C)}\right)\left(\sigma^{(b)}+\epsilon^{(C)}\right)
    }
\end{equation}
where $\langle\cdot\rangle$ denotes the temporal mean, $\sigma^{(a)}$ and $\sigma^{(b)}$ are the temporal standard deviations of $a(t)$ and $b(t)$, and $\epsilon^{(C)}$ is a small positive constant used for numerical stability. The correlation ranges from $-1$ (anti-aligned) to $1$ (perfectly aligned). In our case, it is used as 
\begin{align}
    \mathcal{L}^{(\text{corr})}
    &= \left(1-C(\bar{x}(t),\bar{y}(t+\ell))\right)
    + \left(1-C(\Delta\bar{x}(t),\Delta\bar{y}(t+\ell))\right) \nonumber\\
    &\quad + \frac{1}{2}\left(1-C(\bar{x}(t)-\langle\bar{x}\rangle,\Delta\bar{y}(t))\right)
    + \frac{1}{2}\left(1-C(\bar{y}(t)-\langle\bar{y}\rangle,-\Delta\bar{x}(t))\right)
\end{align}
where $\bar{x}(t)$ and $\bar{y}(t)$ are the smoothed versions of the normalised sheep and wolf population time series, respectively, with
\begin{equation}
    x(t)=\frac{N^{(\mathcal{S})}(t)}{N^{(\mathcal{S})}_{\max}},
    \qquad
    y(t)=\frac{N^{(\mathcal{W})}(t)}{N^{(\mathcal{W})}_{\max}}.
\end{equation}
The smoothing is performed using a uniform moving-average kernel of width $m$. Further, $\Delta\bar{x}(t)=\bar{x}(t+\Delta t)-\bar{x}(t)$, and similarly for $\Delta\bar{y}(t)$. We also compare the sheep population at time $t$ with the wolf population shifted by a delay $\ell$. The rationale is that in a classical Lotka-Volterra curve, the sheep population leads the wolf population, because an increase in prey availability supports predator growth only after a delay, while the subsequent increase in predators suppresses the prey population. Thus, the first two terms in $\mathcal{L}^{(\text{corr})}$ reward the lagged alignment of prey and predator population levels and their slopes. The last two terms encode the local Lotka-Volterra pressure, namely that above-average sheep abundance should be associated with wolf growth, while above-average wolf abundance should be associated with sheep decline.

Other components of the main loss function discourage degenerate solutions that satisfy the correlation loss $\mathcal{L}^{(\text{corr})}$ without showing sustained Lotka-Volterra-like oscillations. These include the extinction loss $\mathcal{L}^{(e)}$ and the ceiling loss $\mathcal{L}^{(c)}$, which penalise population trajectories going below and above prescribed thresholds, respectively. The former prevents population collapse, while the latter prevents saturation at the artificial population capacity of the simulation. Next, the amplitude loss $\mathcal{L}^{(a)}$ requires both populations to show sufficiently large oscillations in both halves of the rollout. The turning loss $\mathcal{L}^{(t)}$ requires both upward and downward movement in each half of the trajectory, thereby rejecting monotonic growth or collapse. The crossing loss $\mathcal{L}^{(\mathrm{cross})}$ penalises trajectories that cross their temporal mean fewer than $n^{(\mathrm{cross})}$ times. Finally, the drift loss $\mathcal{L}^{(d)}$ penalises long-term changes in the mean population level by comparing the beginning and end of the rollout. %Finally, the terminal edge loss $\mathcal{L}^{(T)}$ applies boundary penalties again on the final part of the trajectory, discouraging solutions that appear oscillatory early but approach extinction or saturation near the end.

Thus, the final loss function for tuning the ABM is a weighted sum of all the loss components described above,
\begin{equation}\label{eq:abm_loss}
    \mathcal{L}
    =
    \sum_{r\in\mathcal{C}}
    \lambda^{(r)}\mathcal{L}^{(r)},
    \qquad
    \mathcal{C}=\{\text{corr},e,c,a,t,\mathrm{cross},d\}
\end{equation}
where each $\lambda^{(\cdot)}$ is a positive weighting constant.

With the agent controllers fixed, we used covariance matrix adaptation
evolutionary strategy (CMA-ES)~\citep{hansen2016cma} to minimise Equation~\eqref{eq:abm_loss} over ABM-level parameters only. The ABM parameter vector subjected to optimisation is given by
\begin{equation}
    \phi^{(\mathrm{ABM})}
    =
    \left[
    k^{(G)}, k^{(P)}, \mu^{(\mathcal{S})}, \mu^{(\mathcal{W})},
    e_b^{(\mathcal{S})}, e_d^{(\mathcal{S})},
    t_b^{(\mathcal{S})}, t_d^{(\mathcal{S})},
    e_b^{(\mathcal{W})}, e_d^{(\mathcal{W})},
    t_b^{(\mathcal{W})}, t_d^{(\mathcal{W})}
    \right],
\end{equation}
where $k^{(G)}$ is the grass intake rate, $k^{(P)}$ is the predation transfer rate, $\mu^{(\mathcal{S})}$ and $\mu^{(\mathcal{W})}$ are the metabolic costs of sheep and wolves, and $e_b,e_d,t_b,t_d$ denote the energy and time thresholds for birth and death for each species. In practice, CMA-ES samples $M$ raw candidate vectors at each generation. Each raw vector is transformed with a sigmoid function and rescaled to a predefined viable range before being evaluated in the ABM. For every candidate vector, one ABM rollout is simulated for $T^{(\mathrm{ABM})}$ steps, after which the resulting sheep and wolf population trajectories are assigned a loss according to Equation~\eqref{eq:abm_loss}. CMA-ES then updates its search distribution using these losses and samples a new set of candidate vectors. This process is repeated for $G^{(\mathrm{ABM})}$ generations.

\subsection{Sensor model}

Agents emit $r$ equiangular rays from $\mathbf{q}_i(t)$, covering the full angular range with maximum length $d_{\mathcal{R}}$. Ray-agent intersections are computed using the standard ray-circle quadratic test~\citep{ericson2004real}. If multiple active agents are intersected, only the closest intersection is retained, making sensing occlusion-aware and producing blind spots (Fig.~\ref{fig1}b).

Each ray $p\in\{1,\dots,r\}$ returns two channels of information. The first channel is the distance from the ray origin to the closest intercepted agent surface. The second channel is a type value indicating whether the intercepted agent is a sheep or a wolf. Let $\tau_j\in\{1,-1\}$ denote the type value of an agent $j$, with sheep encoded as $1$ and wolves encoded as $-1$. The observation of ray $p$ emitted by agent $i$ is given by
\begin{equation}
    \mathbf{o}_{i,p}(t)
    =
    \begin{cases}
        \left[d_{i,p}(t),\tau_j\right]^\top,
        & \text{if ray } p \text{ first intersects active agent } j,\\
        \left[d_{\mathcal{R}},0\right]^\top,
        & \text{if no active agent is intersected.}
    \end{cases}
\end{equation}
Here, $d_{i,p}(t)$ is the distance along the ray to the first valid intersection, and $\tau_{j}(t)$ is the type of agent $j$ intersected by the ray. The external observation vector is obtained by concatenating all ray observations,
\begin{equation}
    \mathbf{o}^{(\mathrm{ex})}_i(t)
    =
    \left[
    d_{i,1}(t),\tau_{i,1}(t),
    d_{i,2}(t),\tau_{i,2}(t),
    \dots,
    d_{i,r}(t),\tau_{i,r}(t)
    \right]^\top .
\end{equation}
The agent controllers also receive a subset of their internal states as input to make them aware of the proprioceptive variables. Hence, the ray observations are concatenated with some internal variables. Let $\eta_i(t)$ denote the pre-metabolic energy gain or loss of agent $i$, and let $f_i(t)\in\{0,1\}$ denote whether agent $i$ is overlapping with at least one other active agent. The internal observation vector is
\begin{equation}
    \mathbf{o}^{(\mathrm{in})}_i(t)
    =
    \left[
    e_i(t),\eta_i(t),f_i(t),
    \|\mathbf{q}_i(t)\|_2,
    \mathbf{v}_i^\top(t),
    \omega_i(t),
    \theta_i(t)
    \right]^\top .
\end{equation}
Thus, the complete observation vector provided to the controller is
\begin{equation}
    \mathbf{o}_i(t)
    =
    \left[
    \mathbf{o}^{(\mathrm{ex})}_i(t),
    \mathbf{o}^{(\mathrm{in})}_i(t)
    \right]^\top .
\end{equation}
%Table: sensor constants: $r$, $d_{\mathcal{R}}$
%Table: ray output channels: $d_{i,p}(t)$, $\tau_{i,p}(t)$
%Table: type encoding: sheep $\tau=1$, wolf $\tau=-1$, empty ray $\tau=0$
%Table: internal observation variables: $e_i(t)$, $\eta_i(t)$, $f_i(t)$, $\|\mathbf{q}_i(t)\|_2$, $\mathbf{v}_i(t)$, $\omega_i(t)$, $\theta_i(t)$
%Table: observation dimensions: $\mathbf{o}^{(\mathrm{ex})}_i(t)\in\mathbb{R}^{2r}$, $\mathbf{o}^{(\mathrm{in})}_i(t)\in\mathbb{R}^{8}$, $\mathbf{o}_i(t)\in\mathbb{R}^{2r+8}$

\subsection{Controller model}
The agents control their motion using velocity controllers, which are modelled as continuous-time recurrent neural networks (CTRNNs). Specifically, we use a modern interpretation of the Wilson-Cowan model~\citep{sussillo2014neural}. For an agent $i$, the hidden state dynamics are given by
\begin{align}\label{eq:controller_update}
    &\dot{\mathbf{z}}_i(t)
    =
    \left(
    \tanh\!\left(
    J\mathbf{z}_i(t) + E\mathbf{o}_i(t) + \mathbf{b}
    \right)
    -
    \mathbf{z}_i(t)
    \right)
    \odot
    \left(c_{\tau}\sigma(\boldsymbol{\tau})\right), \\
    &\left[ a_i^{(1)}(t), \,a_i^{(2)}(t)\right]^{\top}
    =
    D\mathbf{z}_i(t).
\end{align}
Here $\mathbf{z}_i(t)\in\mathbb{R}^{h}$ is the controller hidden state, interpreted as neuronal population activity~\citep{sussillo2014neural}. The recurrent matrix is $J\in\mathbb{R}^{h\times h}$, the observation matrix is $E\in\mathbb{R}^{h\times(2r+8)}$, $\mathbf{b}\in\mathbb{R}^{h}$ is a bias, $\boldsymbol{\tau}\in\mathbb{R}^{h}$ contains learnable time-scale parameters, and $D\in\mathbb{R}^{2\times h}$ maps hidden states to readouts. The controller learnable parameter set is $\phi^{(\mathcal{A})}=\{J,E,\mathbf{b},\boldsymbol{\tau},D\}$.

%\subsection{Agent controller optimization}
%To obtain an optimised agent controller parameter $\phi^{{\mathcal{A}*}}$, we train a common wolf and a common sheep controller using the same-rollout evaluation trick used in~\citep{chaturvedi2025emergence}. First, to train the wolves we sample different parameter vectors $\phi^{\mathcal{A}}_i, \forall i\in \mathcal{W}$ for their controller from a CMA-ES distribution. Then, we evaluate all the wolves against randomly initialized sheep brains $\phi^{\mathcal{A}}_j\sim U(-1,1), \forall j\in \mathcal{S}$. The fitness for each wolf at the end of rollout is given as $f_i = e_i(T) - e_i(0)$. The evaluated fitness is averaged across $H$ scenarios for each CMA-ES sample. Finally, the CMA-ES algorithm optimizes the wolf controller parameter over $G^{(\mathcal{W})}$ generations. During this optimization the number of agents are held constant $N^{\mathcal{W}}(t) = N^{\mathcal{W}}_{\max}$ and $N^{\mathcal{S}}(t) = N^{\mathcal{S}}_{\max}$. The controller for sheep agent is optimised in a similar way by training them against random wolf agents. We use random opponents to avoid overfitting to a single adversarial strategy.

\subsection{Agent controller optimisation} \label{subsec:agent_optim}
Before ABM-level tuning, we optimise separate sheep and wolf controller parameters, \(\phi^{(\mathcal{S})*}\) and \(\phi^{(\mathcal{W})*}\), using the same-rollout evaluation trick used in~\citep{chaturvedi2025emergence}. First, to train the wolves, we sample different controller parameter vectors $\phi^{(\mathcal{W})}_i$, for all $i\in \mathcal{W}$, from a CMA-ES distribution. We then evaluate all wolves against sheep whose controller parameters are randomly initialised as $\phi^{(\mathcal{S})}_j\sim U(-1,1)$, for all $j\in \mathcal{S}$. The fitness for each wolf at the end of a rollout is given as $f_i = e_i(T) - e_i(0)$. The fitness assigned to each CMA-ES sample is averaged across $H$ scenarios having different initial conditions and ABM parameters $\phi^{(\mathrm{ABM})}$. Finally, CMA-ES optimises the wolf controller parameters over $G^{(\mathcal{W})}$ generations.

During this optimisation phase, the number of agents is held constant, such that $N^{(\mathcal{W})}(t)=N^{(\mathcal{W})}_{\max}$ and $N^{(\mathcal{S})}(t)=N^{(\mathcal{S})}_{\max}$. The sheep controller is optimised in the same way by training sheep against randomly initialised wolf controllers. We use random opponents to reduce overfitting to a single adversarial strategy. After optimisation, the mean controller parameters of the respective CMA-ES distributions are used as the fixed sheep and wolf controllers in the Lotka-Volterra tuning phase.

%Table: controller dimensions: $h$, $2r+8$, $2$
%Table: controller state variables: $\mathbf{z}_i(t)$, $a_i^{(1)}(t)$, $a_i^{(2)}(t)$
%Table: controller parameters: $J$, $E$, $\mathbf{b}$, $\boldsymbol{\tau}$, $D$
%Table: controller constants: $c_{\tau}$
%Table: controller parameter vector: $\phi^{(\mathcal{A})}=\{J,E,\mathbf{b},\boldsymbol{\tau},D\}$
%Table: controller initialization: $\mathbf{z}_i(0)$, initial action values
%Table: agent training settings: CMA-ES population size, elite ratio, initial step size, $H$, $G^{(\mathcal{W})}$, $G^{(\mathcal{S})}$
%Table: opponent initialization ranges: sheep random controller range $U(-1,1)$, wolf random controller range $U(-1,1)$
%Table: agent training rollout settings: fixed $N^{(\mathcal{S})}_{\max}$, fixed $N^{(\mathcal{W})}_{\max}$, rollout length for sheep training, rollout length for wolf training
%Table: agent training objective: $f_i=e_i(T)-e_i(0)$
%Table: optimised controller outputs: $\phi^{(\mathcal{S})*}$, $\phi^{(\mathcal{W})*}$

\subsection{Simulation details}

The entire model was implemented using the \texttt{JAX}~\cite{bradbury2018jax} based \texttt{ABMax} framework~\citep{chaturvedi2025abmax}, which enables batched vectorisation across multiple instances of the model. Specifically, the rank-match update algorithm from the framework allowed different model instances to maintain different numbers of active agents while being evaluated in parallel. We adopted the CMA-ES algorithm implemented in the \texttt{Evosax} library~\citep{lange2023evosax}. The values of all parameters and initial states are summarised in Table~\ref{tab:parameters}. Unless otherwise stated, these values are used for all simulations. All simulations were run on an NVIDIA A100 GPU. The code for the model is available online.\footnote{\texttt{https://github.com/corimandl/LV\_ABM.git}}

\begin{table}[t]
\centering
\tiny
\setlength{\tabcolsep}{3pt}
\renewcommand{\arraystretch}{1.05}
\caption{Model parameters and initial conditions}
\label{tab:parameters}
\begin{tabular*}{\textwidth}{@{\extracolsep{\fill}}ll ll ll}
\hline
\textbf{Symbol} & \textbf{Value} &
\textbf{Symbol} & \textbf{Value} &
\textbf{Symbol} & \textbf{Value} \\
\hline

\multicolumn{6}{l}{\textbf{ABM and initial conditions}}\\
$N^{(\mathcal{S})}_{\max}$ & $400$ &
$N^{(\mathcal{W})}_{\max}$ & $400$ &
$N^{(\mathcal{S})}(0)$ & $150$ \\
$N^{(\mathcal{W})}(0)$ & $80$ &
$\Delta t$ & $0.1$ &
$(q_x^{\max},q_y^{\max})$ & $(500,500)$ \\
$R^{(\mathcal{S})}$ & $5$ &
$R^{(\mathcal{W})}$ & $5$ &
$d_G$ & $200$ \\
$d_s$ & $30$ &
$e_{\max}$ & $100$ &
$\mathbf{q}_i(0)$ & $\mathcal{U}([-250,250]^2)$ \\
$\mathbf{v}_i(0)$ & $\mathbf{0}$ &
$\theta_i(0)$ & $\mathcal{U}[-\pi,\pi)$ &
$\omega_i(0)$ & $0$ \\
$\epsilon$ & $0.05$ &
$k^{(s)}$ & $2R/\Delta t$ &
$k^{(u)}$ & $2.0$ \\
\hline

\multicolumn{6}{l}{\textbf{Lotka--Volterra objective}}\\
$m$ & $51$ &
$\ell$ & $600$ &
$\epsilon^{(C)}$ & $10^{-8}$ \\
$(\rho_{\mathrm{low}},\rho_{\mathrm{up}},\rho_{\mathrm{ceil}})$ & $(0.05,0.88,0.97)$ &
$A_{\min}$ & $0.20$ &
$\Delta_{\min}$ & $0.12$ \\
$n^{(\mathrm{cross})}$ & $3$ &
$\delta_{\mathrm{drift}}$ & $0.12$ &
$\lambda^{(\phi)}$ & $35$ \\
$\lambda^{(e)}$ & $1200$ &
$\lambda^{(c)}$ & $1000$ &
$\lambda^{(a)}$ & $500$ \\
$\lambda^{(t)}$ & $800$ &
$\lambda^{(\mathrm{cross})}$ & $50$ &
$\lambda^{(d)}$ & $500$ \\
$T^{(\mathrm{ABM})}$ & $6000$ &
 &  &
 &  \\
\hline

\multicolumn{6}{l}{\textbf{Optimised ABM parameter ranges}}\\
$k^{(G)}$ & $[0.02,0.50]$ &
$k^{(P)}$ & $[0.45,0.90]$ &
$\mu^{(\mathcal{S})}$ & $[0.006,0.090]$ \\
$\mu^{(\mathcal{W})}$ & $[0.008,0.090]$ &
$e_b^{(\mathcal{S})}$ & $[0.30,0.60]e_{\max}$ &
$e_d^{(\mathcal{S})}$ & $[0.005,0.10]e_{\max}$ \\
$t_b^{(\mathcal{S})}$ & $[0.003,0.020]T$ &
$t_d^{(\mathcal{S})}$ & $[0.001,0.040]T$ &
$e_b^{(\mathcal{W})}$ & $[0.30,0.60]e_{\max}$ \\
$e_d^{(\mathcal{W})}$ & $[0.005,0.10]e_{\max}$ &
$t_b^{(\mathcal{W})}$ & $[0.003,0.040]T$ &
$t_d^{(\mathcal{W})}$ & $[0.001,0.020]T$ \\
\hline

\multicolumn{6}{l}{\textbf{Sensing and controller}}\\
$r$ & $13$ &
$d_{\mathcal{R}}^{(\mathcal{S})}$ & $300$ &
$d_{\mathcal{R}}^{(\mathcal{W})}$ & $300$ \\
$\tau$ & $\{-1,0,1\}$ &
$\mathbf{o}_i(t)$ & $\mathbb{R}^{2r+8}$ &
$h$ & $60$ \\
$c_{\tau}$ & $10$ &
$\phi^{(\mathcal{A})}$ & $\{J,E,\mathbf{b},\boldsymbol{\tau},D\}$ &
$\mathbf{z}_i(0),\mathbf{a}_i(0)$ & $\mathbf{0},\mathbf{0}$ \\
\hline

\multicolumn{6}{l}{\textbf{CMA-ES settings}}\\
$M$ & $16$ &
$G^{(\mathrm{ABM})}$ & $150$ &
$M^{(\mathcal{S})}$ & $50$ \\
$M^{(\mathcal{W})}$ & $50$ &
$G^{(\mathcal{S})}$ & $4000$ &
$G^{(\mathcal{W})}$ & $4000$ \\
$H$ & $12$ &
$T^{(\mathcal{S})},T^{(\mathcal{W})}$ & $1000,1000$ &
$\phi^{(\mathcal{S})}_{\mathrm{rand}}$ & $\mathcal{U}(-2,2)$ \\
$\phi^{(\mathcal{W})}_{\mathrm{rand}}$ & $\mathcal{U}(-1,1)$ &
$f_i$ & $e_i(T)-e_i(0)$ &
$\alpha_{\mathrm{elite}},\sigma_0$ & default \\
\hline
\end{tabular*}
\end{table}

% LV curve, constant derivation
\section{Results}
\begin{figure}
\centering
\includegraphics[width=1.0\textwidth]{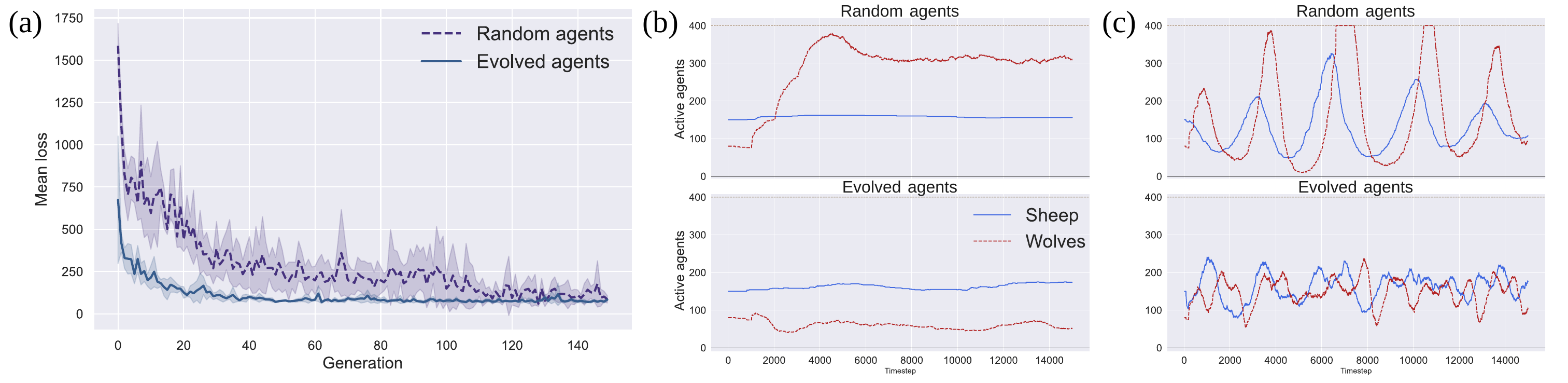}
\caption{Lotka-Volterra tuning in the predator-prey ABM. \textbf{(a)} ABM-level loss during optimisation of \(\phi^{(\mathrm{ABM})}\) with fixed random or evolved controllers. Lines show means across four seeds and bands show one standard deviation. \textbf{(b)} Population trajectories under random ABM parameters. \textbf{(c)} Population trajectories after ABM-parameter optimisation.}
\label{fig:results1}
\end{figure}

 The results of the ABM-level parameter estimation can be seen in Fig.~\ref{fig:results1}a. During this stage, agent controllers are fixed and only the ABM parameters \(\phi^{(\mathrm{ABM})}\) are optimised. We optimise for two fixed-controller settings: random controllers \(\phi^{(\mathcal{A})}\sim U(-1,1)\), and evolved controllers \(\phi^{(\mathcal{S})*},\phi^{(\mathcal{W})*}\) obtained from the controller-optimisation stage. The loss decreases in both settings, indicating that the proposed objective can guide ABM parameters toward lower-loss regimes. However, the loss decreases faster and converges to a lower value for evolved controllers. To make optimisation possible for random controllers, the grass patch spans the entire arena, so that randomly moving sheep have a constant energy supply.

Fig.~\ref{fig:results1}b shows the population trajectories over a rollout of \(T=15000\) steps using randomly sampled ABM parameters. In both random and evolved controller settings, the resulting trajectories miss key characteristics of Lotka-Volterra dynamics, in particular sustained coupled oscillations between sheep and wolves. After optimisation using the procedure in Section~\ref{subsec:lv_objective}, the population trajectories acquire oscillatory predator-prey structure and remain active for the duration of the rollout (Fig.~\ref{fig:results1}c).

Interestingly, the optimised random-agent setting produces smoother and more regular oscillations in the upper panel of Fig.~\ref{fig:results1}c, but these oscillations also approach the artificial population ceiling. In contrast, the evolved-agent setting produces noisier and less idealised oscillations, but the trajectories remain more bounded and avoid the most extreme saturation events. This suggests that the loss can induce LV-like macroscopic structure in both settings, while learned controllers produce a more behaviourally grounded and less idealised population dynamics. This distinction is consistent with the fact that the classical Lotka-Volterra model is highly idealised and omits several ecological features present in spatial, individual-based predator-prey systems, such as finite resources, spatial structure, behavioural heterogeneity, and individual-level stochasticity \citep{wangersky1978lotka,raz2017volterra,muhlbauer2020gauser}.

\begin{figure}
\centering
\includegraphics[width=1.0\textwidth]{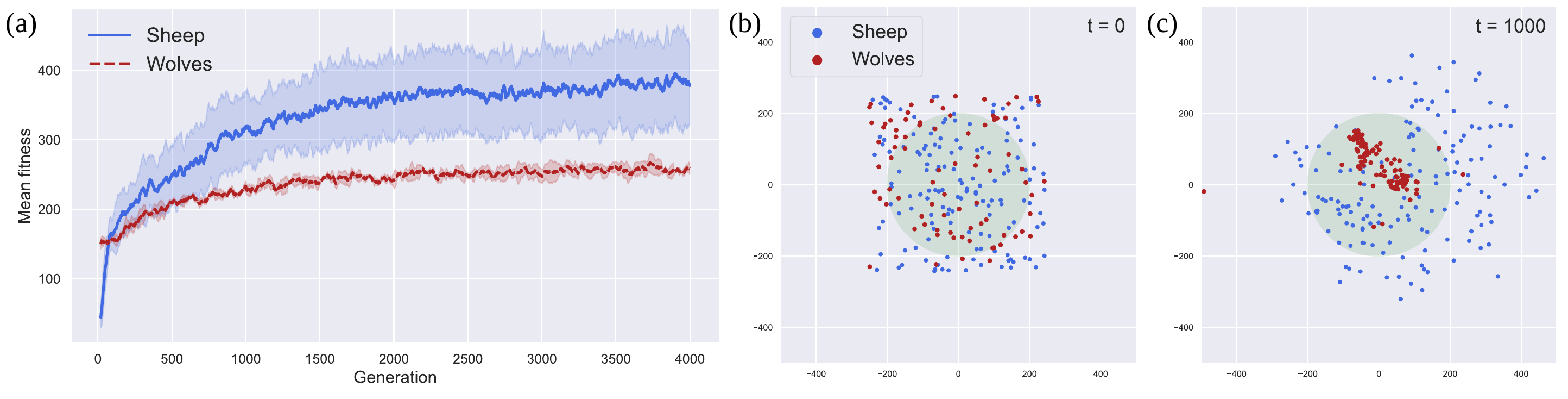}
\caption{Agent controller optimisation and spatial organisation. \textbf{(a)} Sheep and wolf fitness curves, shown as mean across three seeds with one-standard-deviation bands. \textbf{(b)} Optimised ABM rollout with evolved agents at $t=0$. \textbf{(c)} Same rollout at $t=1000$, showing wolf aggregation near the grass patch and broader sheep dispersion.}
 \label{fig:results2}
\end{figure}

Next, we rendered the optimised ABM using the evolved controllers from the separate controller-optimisation stage (Fig.~\ref{fig:results2}a). The snapshots in Fig.~\ref{fig:results2}b and c show that learned controller dynamics induce clear spatial organisation. Wolves aggregate in and around the grass patch, where sheep density is higher, while sheep remain more broadly dispersed. This behaviour was already present before ABM-parameter optimisation, suggesting that ABM tuning mainly makes this behavioural ecology viable over longer rollouts by balancing energy intake, predation, reproduction, and death. A video is available online.\footnote{\url{https://youtu.be/vBS8EfF8GZ0}}

% behaviour
% ablation (random sheep, random wolves)

\section{Discussion}

In this work, we showed that a continuous agent-based predator-prey model can be tuned such that its population trajectories show signatures of Lotka-Volterra-like dynamics. The implication is not that the ABM discovers Lotka-Volterra dynamics automatically, nor that a unique microscopic model can be inferred from a macroscopic trajectory. Rather, the result shows that a desired macroscopic ecological regime can be used as a top-down constraint for searching in the ABM parameter space. In this sense, Lotka-Volterra dynamics are used not as a replacement for the ABM, but as a target that identifies parameter regimes in which the ABM remains dynamically well behaved. This also demonstrates a use case for \texttt{ABMax}, where many ABM instances with different numbers of active agents can be evaluated in parallel on hardware accelerators.

The results further suggest a distinction between behavioural structure and population-level stabilisation. The evolved controllers produced spatial predator-prey behaviour, with wolves aggregating near the grass patch and sheep spreading more broadly. This behaviour was not created by the Lotka-Volterra loss itself. Rather, the ABM-level tuning made the consequences of such behaviour viable over longer horizons by balancing energy intake, predation, reproduction, and death. Thus, our goal is different from showing that predator and prey agents optimised only for local behaviour necessarily produce Lotka-Volterra dynamics. Instead, we ask whether, once local behaviour exists, ecological parameters can be tuned so that the resulting population dynamics satisfy a desired macroscopic constraint. A limitation is that the loss function is indirect and hand designed. However, a direct mean-squared error to a single Lotka-Volterra trajectory would also be difficult to justify because exact trajectory matching is unrealistic.

Future work can make the model more ecological and adaptive. For instance, internal energy could affect mass, speed, turning radius, or metabolic cost, and the grass patch could become a depleted and regenerating resource rather than a fixed energy source. Another direction is to allow predator and prey controllers to adapt online, creating a moving target for both populations. More broadly, this work points toward a way of studying open-ended multi-agent systems through ecological constraints. Instead of allowing rich agent interactions to drift without structure, macroscopic ecological targets may help keep the system within viable bounds while still preserving non-trivial dynamics.

\begin{credits}
\subsubsection{\ackname} 
This publication is part of the project Dutch Brain Interface Initiative (DBI$^2$) with project number 024.005.022 of the research programme Gravitation which is (partly) financed by the Dutch Research Council (NWO).

\subsubsection{\discintname}
The authors have no competing interests to declare that are
relevant to the content of this article.
\end{credits}
\par\bigskip


\begingroup
  \let\clearpage\relax
  \let\newpage\relax
  
\bibliographystyle{splncs04}
\bibliography{samplepaper}

@article{holland1992complex,
  title = {Complex Adaptive Systems},
  author = {Holland, John H.},
  journal = {Daedalus},
  volume = {121},
  number = {1},
  pages = {17--30},
  year = {1992},
  publisher = {MIT Press}
}

@article{deangelis2005individual,
  title = {Individual-Based Modeling of Ecological and Evolutionary Processes},
  author = {DeAngelis, Donald L. and Mooij, Wolf M.},
  journal = {Annual Review Of Ecology, Evolution, And Systematics},
  volume = {36},
  pages = {147--168},
  year = {2005},
  doi = {10.1146/annurev.ecolsys.36.102003.152644}
}

@article{volterra1926fluctuations,
  title = {Fluctuations in the Abundance of a Species Considered Mathematically},
  author = {Volterra, Vito},
  journal = {Nature},
  volume = {118},
  number = {2972},
  pages = {558--560},
  year = {1926},
  doi = {10.1038/118558a0}
}

@phdthesis{idema2005behaviour,
  title = {The Behaviour and Attractiveness of the {L}otka-{V}olterra Equations},
  author = {Idema, Timon},
  school = {Universiteit Leiden},
  year = {2005}
}

@article{yaya2023predator,
  title = {A Predator--Prey Model from a Collective Dynamics and Self-Propelled Particles Approach},
  author = {Yaya, Youssouf Yaya},
  journal = {Computer Sciences \& Mathematics Forum},
  volume = {7},
  number = {1},
  pages = {50},
  year = {2023},
  publisher = {MDPI},
  doi = {10.3390/iocma2023-14375}
}

@misc{chaturvedi2025abmax,
  title = {Abmax: A JAX-Based Agent-Based Modeling Framework},
  author = {Chaturvedi, Siddharth and El-Gazzar, Ahmed and van Gerven, Marcel},
  year = {2025},
  eprint = {2508.16508},
  archivePrefix = {ArXiv}
}

@misc{bradbury2018jax,
  title = {JAX: Composable Transformations of Python+NumPy Programs},
  author = {Bradbury, James and Frostig, Roy and Hawkins, Peter and Johnson, Matthew James and Leary, Chris and Maclaurin, Dougal and Necula, George and Paszke, Adam and VanderPlas, Jake and Wanderman-Milne, Skye and Zhang, Qiao},
  year = {2018},
  howpublished = {\url{http://github.com/jax-ml/jax}}
}

@incollection{helbing2012agent,
  title={Agent-based Modeling},
  author={Helbing, Dirk},
  booktitle={Social self-organization: Agent-based simulations and experiments to study emergent social behavior},
  pages={25--70},
  year={2012},
  publisher={Springer}
}

@article{clauset2009power,
  author  = {Clauset, Aaron and Shalizi, Cosma Rohilla and Newman, M. E. J.},
  title   = {Power-Law Distributions in Empirical Data},
  journal = {SIAM Review},
  volume  = {51},
  number  = {4},
  pages   = {661--703},
  year    = {2009},
  doi     = {10.1137/070710111}
}

@article{quan2026macroscopic,
  author  = {Quan, Quan and Yu, Xinchen and Li, Yue and Qi, Guoyuan},
  title   = {Macroscopic modelling and analysis based on microscopic models for swarm systems},
  journal = {Scientific Reports},
  volume  = {16},
  pages   = {10342},
  year    = {2026},
  doi     = {10.1038/s41598-026-38163-w}
}

@article{monti2023learning,
  author  = {Monti, Corrado and Pangallo, Marco and De Francisci Morales, Gianmarco and Bonchi, Francesco},
  title   = {On learning agent-based models from data},
  journal = {Scientific Reports},
  volume  = {13},
  pages   = {9268},
  year    = {2023},
  doi     = {10.1038/s41598-023-35536-3}
}

@article{niemann2021datadriven,
  author  = {Niemann, Jan-Hendrik and Klus, Stefan and Sch{\"u}tte, Christof},
  title   = {Data-driven model reduction of agent-based systems using the Koopman generator},
  journal = {PLOS ONE},
  volume  = {16},
  number  = {5},
  pages   = {e0250970},
  year    = {2021},
  doi     = {10.1371/journal.pone.0250970}
}

@article{hodzic2016complex,
  author  = {Hodzic, Migdat and Selman, Suvad and Hadzikadic, Mirsad},
  title   = {Complex ecological system modeling},
  journal = {Periodicals Of Engineering And Natural Sciences},
  volume  = {4},
  number  = {1},
  pages   = {44--50},
  year    = {2016},
  doi     = {10.21533/pen.v4i1.9}
}

@article{chaturvedi2025emergence,
  title={Emergence of Internal State-Modulated Swarming in Multi-Agent Patch Foraging System},
  author={Chaturvedi, Siddharth and El-Gazzar, Ahmed and van Gerven, Marcel},
  journal={ArXiv preprint arXiv:2510.18886},
  year={2025}
}

@article{chaturvedi2026role,
  title={Role Differentiation in a Coupled Resource Ecology under Multi-Level Selection},
  author={Chaturvedi, Siddharth and El-Gazzar, Ahmed and van Gerven, Marcel},
  journal={ArXiv preprint arXiv:2604.00810},
  year={2026}
}

@misc{Wilensky1997,
  author       = {Wilensky, Uri},
  title        = {NetLogo Wolf Sheep Predation model},
  year         = {1997},
  institution  = {Center for Connected Learning and Computer‐Based Modeling, Northwestern University},
  address      = {Evanston, IL},
  url          = {http://ccl.northwestern.edu/netlogo/models/WolfSheepPredation}
}

@book{ericson2004real,
  title={Real-Time Collision Detection},
  author={Ericson, Christer},
  year={2004},
  publisher={Crc Press}
}

@article{sussillo2014neural,
  title={Neural Circuits as Computational Dynamical Systems},
  author={Sussillo, David},
  journal={Current Opinion In Neurobiology},
  volume={25},
  pages={156--163},
  year={2014},
  publisher={Elsevier}
}

@article{lee1988thirteen,
  title={Thirteen ways to look at the correlation coefficient},
  author={Lee Rodgers, Joseph and Nicewander, W Alan},
  journal={The American Statistician},
  volume={42},
  number={1},
  pages={59--66},
  year={1988},
  publisher={Taylor \& Francis}
}

@article{hansen2016cma,
  title={The CMA Evolution Strategy: A Tutorial},
  author={Hansen, Nikolaus},
  journal={ArXiv preprint arXiv:1604.00772},
  year={2016}
}

@inproceedings{lange2023evosax,
  title={Evosax: Jax-Based Evolution Strategies},
  author={Lange, Robert Tjarko},
  booktitle={Proceedings of the Companion Conference On Genetic And Evolutionary Computation},
  pages={659--662},
  }

@article{wangersky1978lotka,
  author  = {Wangersky, Peter J.},
  title   = {{L}otka-{V}olterra Population Models},
  journal = {Annual Review Of Ecology And Systematics},
  volume  = {9},
  pages   = {189--218},
  year    = {1978},
  doi     = {10.1146/annurev.es.09.110178.001201}
}

@article{raz2017volterra,
  author  = {R{\"a}z, Tim},
  title   = {The Volterra Principle Generalized},
  journal = {Philosophy Of Science},
  volume  = {84},
  number  = {2},
  pages   = {209--229},
  year    = {2017},
  doi     = {10.1086/690717}
}

@article{muhlbauer2020gauser,
  author  = {M{\"u}hlbauer, Lukas K. and Schulze, A. D. and Harpole, W. Stanley and Clark, Adam T.},
  title   = {gauseR: Simple methods for fitting {L}otka-{V}olterra models describing {G}ause's “Struggle for Existence”},
  journal = {Ecology And Evolution},
  volume  = {10},
  number  = {24},
  pages   = {13275--13283},
  year    = {2020},
  doi     = {10.1002/ece3.6926}
}
\endgroup

\end{document}